\documentclass{article}
\usepackage{arxiv}
\usepackage[utf8]{inputenc} % allow utf-8 input
\usepackage[T1]{fontenc}    % use 8-bit T1 fonts
\usepackage{hyperref}       % hyperlinks
\usepackage{url}            % simple URL typesetting
\usepackage{booktabs}       % professional-quality tables
\usepackage{amsfonts}       % blackboard math symbols
\usepackage{nicefrac}       % compact symbols for 1/2, etc.
\usepackage{microtype}      % microtypography
\usepackage{lipsum}
\usepackage{lineno,hyperref}
\usepackage{algorithm,algpseudocode}
\usepackage{amsmath}
\usepackage{algorithmicx}
\usepackage{verbatim}
\usepackage{blindtext}
\usepackage{algpseudocode}
\usepackage{acro}
\usepackage{doi}
\usepackage[normalem]{ulem}
\usepackage{graphicx}
\usepackage{epstopdf}
\usepackage{subcaption}
\usepackage{tabularx}
\usepackage{array}
\usepackage{float}
\usepackage{multirow}
\usepackage{multicol}
\usepackage{rotating}
\usepackage{enumitem}
\usepackage{pifont}
\usepackage{bigstrut}
\usepackage{color,soul}

\title{Hybrid quantum convolutional neural networks model for COVID-19 prediction using chest X-Ray images}

\author{
 Essam H.~Houssein\\
 Faculty of Computers and Information\\ 
 Minia University \\
 Egypt\\
 \texttt{essam.halim@mu.edu.eg} \\
   \And
  Zainab Abohashima \\
  Faculty of Computer Science\\
  Nahda University in Beni-Suef\\
  Egypt\\
  \texttt{zeinababohashima20@gmail.com} \\
  \And
  Mohamed Elhoseny \\
Department of Computer Science\\ 
American University in the Emirates\\
 Dubai, UAE.\\
  \texttt{melhoseny@ieee.org} \\
  \And
 Waleed M.~Mohamed \\
  Faculty of Computers and Information\\ 
  Minia University\\ 
  Egypt\\
  \texttt{waleedmakram@mu.edu.eg} \\
  }
  
\begin{document}
\maketitle
%\beg{frontmatter}
	\captionsetup[table]{
		labelsep = newline,
		labelfont = bf,
		name = Table,
		justification=justified,
		singlelinecheck=false,%%%%%%% a single line is centered by default
		skip = \medskipamount}	
	\newcolumntype{L}[1]{>{\raggedright\arraybackslash}p{#1}}
	\newcolumntype{C}[1]{>{\centering\arraybackslash}p{#1}}
	\newcolumntype{R}[1]{>{\raggedleft\arraybackslash}p{#1}}

%E-mail: essam.halim@mu.edu.eg
%E-mail: zeinababohashima20@gmail.com
%E-mail: melhoseny@ieee.org
%E-mail: waleedmakram@mu.edu.eg
%E-mail: mohamed\_elhoseny@mans.edu.eg

\begin{abstract}
%% 200 words
Despite the great efforts to find an effective way for COVID-19 prediction, the virus nature and mutation represent a critical challenge to diagnose the covered cases. However, developing a model to predict COVID-19 via Chest X-Ray (CXR) images with accurate performance is necessary to help in early diagnosis. In this paper, a hybrid quantum-classical convolutional Neural Networks (HQCNN) model used the random quantum circuits (RQCs) as a base to detect COVID-19 patients with CXR images. A collection of 6952 CXR images, including 1161 COVID-19, 1575 normal, and 5216 pneumonia images, were used as a dataset in this work. The proposed HQCNN model achieved higher performance with an accuracy of 98.4\% and a sensitivity of 99.3\% on the first dataset cases. Besides, it obtained an accuracy of 99\% and a sensitivity of 99.7\% on the second dataset cases. Also, it achieved accuracy, and sensitivity of 88.6\%, and 88.7\%, respectively, on the third multi-class dataset cases. Furthermore, the HQCNN model outperforms various models in balanced accuracy, precision, F1-measure, and AUC-ROC score. The experimental results are achieved by the proposed model prove its ability in predicting positive COVID-19 cases.
\end{abstract}

%\end{frontmatter}

%\linenumbers

%%%%%%%%%%%%%%%%%%%%%%%%%%%%%%%%%%%%%%%%%%
%\printacronyms[include-classes=abbrev,name=Abbreviations]
\begin{table}[!htb]
	%\centering
	\begin{tabular}{llll}
		\textbf{Abbreviations} & \\
		CNNs &  Convolutional Neural Networks &
		RQCs   & Random Quantum Circuits \\ 
		QC & Quantum Computing &
		COVID-19    &  Coronavirus Disease 2019  \\
		ReLU   & Rectified Linear Unit &
		DL & Deep Learning \\
		Conv  & Convolutional  &
		CXR & Chest X-Ray \\ 
		CT & Computed Tomography
	\end{tabular}
\end{table}
%%%%%%%%%%%%%%%%%%%%%%%%%%%%%%%%%%%%%%%%%%

\section{Introduction}
\label{Sec:Introduction}
Recently, COVID-19 has been rapidly spreading in several countries caused by infection of human beings with severe acute respiratory syndrome coronavirus $2$ (SARS-CoV-2) \cite{li2020coronavirus}. The ongoing COVID-19 pandemic attacks human health causing respiratory disease and acute kidney injury \cite{sise2020case}. COVID-19 stands for coronavirus disease 2019, which is a new type of coronavirus and falls as a subtype of RNA viruses \cite{kooraki2020coronavirus}. The genetic structure of COVID-19 is identical to bat-coronavirus RaTG13, MERS-coronavirus, and SARS-coronavirus by 95\%, 50\%, and 82\% respectively \cite{ahmed2020preliminary,udugama2020diagnosing}. The first outbreak of the disease was in China at the end of 2019. The most common clinical symptoms of the disease are fever, sore throat, cough, dyspnea, and muscle pain \cite{jamil2020diagnosis}. In March 2020, the World Health Organization (WHO) reported COVID-19 as a pandemic \cite{li2020coronavirus}. The most common diagnostic tool used for COVID-19 prediction is a real-time "reverse transcription-polymerase chain reaction" (RT-PCR) \cite{salehi2020coronavirus}, even though RT-PCR has low sensitivity with the early phases \cite{xie2020chest}. Alternative imaging tools (i.e., chest X-ray (CXR) and computed tomography (CT) scans) play an important and critical role in COVID-19 prediction \cite{li2020ct}. Radiologists prefer to use the CXR to diagnose chest diseases. The CXR has been used to detect infected COVID-19 cases \cite{chen2020epidemiological}. Thus utilizing the artificial neural networks to detect COVID-19 in CXR has become needed.

%Recently, many meta-heuristic algorithms have been used to overcome the high processing time and accuracy problems. However, several meta-heuristic algorithms have been proposed in the recent years such as; Archimedes optimization algorithm \cite{hashim2020archimedes}, L{\'e}vy flight distribution \cite{houssein2020levy}, Monarch butterfly optimization \cite{wang2019monarch}, and Henry gas solubility optimization \cite{hashim2019henry}. Since many fields of science need optimization, meta-heuristic algorithms have been applied successfully to solve many optimization problems in different domains, such as drug design \cite{houssein2020hybrid}, Maximizing lifetime of wireless sensor networks \cite{ahmed2019maximizing}, Bioinformatics \cite{hashim2020modified}, Information feedback \cite{zhang2020enhancing}, Job-shop scheduling problem \cite{gao2020solving}, and feature selection \cite{neggaz2020efficient}. In general speaking, several meta-heuristic algorithms have been applied to present the COVID-19 disease classification and prediction, for instance, the marine predators' algorithm (MPA) \cite{abdel2020hybrid} and Salp swarm algorithm (SSA) \cite{al2020optimization}.%

The use of Artificial Neural Networks (ANNs) in medical data has significantly advanced. The CNNs are one of the most powerful, and widespread deep learning models in pattern recognition and image classification. Furthermore, CNNs have achieved superior performance in medical image detection \cite{miller2018artificial}. Various works for the prediction of COVID-19 cases using CNNs structure have been proposed, these works are presented in detail in section "Related Work". Chowdhury et al. \cite{chowdhury2020can} proposed deep CNNs models to pre-trained their model using transfer learning to discuss the leverage of artificial intelligence (AI) for classification COVID-19 on CXR images. Especially that the computational power of classical ANNs can’t learn large training data with low-cost \cite{lundervold2019overview}. In a particular, it is not capable of generating more kernels (kernel estimation) with high-dimensional features \cite{havlivcek2019supervised,schuld2019quantum}. Effortlessly, quantum computing is capable of performing complex kernels in n-dimensional space. 

On the other hand, the quantum computing (QC) field has proven its significant role in intractable problems with classical counterparts via quantum supremacy  \cite{arute2019quantum,boixo2018characterizing} and with the upgrowth of the concept of quantum computing (QC) and its improvements in the machine learning field (i.e., learning capacity, run-time, and learning efficiency) \cite{dunjko2018machine}. Quantum computing has also demonstrated a remarkable influence in machine learning (ML) on near-term quantum computers \cite{killoran2019continuous}. Quantum neural network approaches have been proposed such as \cite{jeswal2019recent,farhi2018classification}. According to our best of knowledge, the start point of QNNs was in 1995. Kak \cite{kak1995quantum} introduced NNs concepts in the quantum computing world. This study focuses on the hybrid “quantum-classical” approach. Henderson et al. \cite{henderson2020quanvolutional} presented a new quantum convolutional layer within CNNs based on quantum circuits to estimate kernel in high-dimensionality. In a similar work \cite{havlivcek2019supervised}, Havlicek et al. proposed a quantum kernel algorithm of support vector machines to estimate kernel with a quantum circuit and to deal with huge features by estimating kernel in n-dimensionality space. Mitarai et al. \cite{mitarai2018quantum} proposed a new hybrid method called QC learning (QCL) based on a quantum circuit. QCL works with large datasets for clustering, regression, and classification tasks.

Therefore, the motivation behind this work is to combine the advantage of the quantum convolutional layer with CNNs and CXR images as an important tool and cheapest diagnostic technique of COVID-19 to produce a new proposed HQCNN model. This model aims to improve the performance of CNNs to detect coronavirus cases in early phases and to speed up the diagnosis and remedy. The HQCNN model is measured with two angle encoding gates and a different number of shots. The better results have been achieved by the Ry angle gate and 1000 shots. The proposed HQCNN model achieved higher performance with an accuracy of 98.4\%, 99\%, and 88.6\% on the first dataset( normal and COVID-19 cases), the second dataset (COVID-19 and pneumonia cases), and the third dataset (normal, COVID-19, and pneumonia cases), respectively. Also, the higher sensitivity and AUC-ROC scores have been achieved with 99.3\%, and 100\%, respectively on the COVID-19 and pneumonia dataset. Besides, the HQCNN model outperforms various models in specificity, balanced accuracy, precision, F1-measure, and AUC-ROC score. The experimental results show the ability of the proposed HQCNN model to classify positive COVID-19 cases. The restrictions of the
HQCNN model are highlighted in Section \ref{Sec:Results}.

The main contributions of this work are:  
\begin{itemize}
	\item A new hybrid CNNs model combined with a quantum circuit called HQCNN for COVID-19 prediction based chest X-ray images is proposed.
	\item The comparative performance investigates of HQCNN model using encoding angle methods and a various number of shots on the quantum device for COVID-19 classification cases.
	\item  HQCNN model is evaluated on binary and multi-class datasets with confirmed COVID-19 cases. 
	\item An exhaustive and a comparative experimental discussion is presented in terms of accuracy, balanced accuracy, sensitivity, specificity, precision, F1-measure, FBeta-measure, AUC-ROC curve, and confusion matrix to evaluate the performance of the proposed HQCNN model.
\end{itemize}
The remainder of this paper is organized as follows: Section \ref{Sec:Related Work} gives an overview of recent COVID-19 classification based the CNNs with CXR and CT images and shows a comparative performance of previous works. Background about the concept of quantum computing (Qubits, and quantum gates) and CNNs are presented in Section \ref{Sec:Preliminaries}. The hybrid proposed HQCNN model is introduced in detail in Section \ref{Sec:Proposed}. Section \ref{Sec:Results} presents the used datasets, performance measures, and the experimental results of the hybrid HQCNN model. Finally, the conclusion of this work and future research are delineated in Section \ref{Sec:Con}.

\section{Related Work}
\label{Sec:Related Work}
This section provides various works of COVID-19 classification based on CNNs that had recently been presented by researchers. Related works are divided into studies based on CXR images and studies based on CT scans. 

In \cite{apostolopoulos2020covid}, a CNNs model (i.e., VGG19, Xception, MobileNetv2) based on transfer learning is proposed to extract meaningful biomarkers of coronavirus patients. Furthermore, the used dataset is collected with 224 COVID-19, 714 pneumonia (bacterial and viral), and 504 normal CXR images to evaluate the performance of CNNs models on the classification of medical images. The results are reported with 96.78\% accuracy, 98.66\% sensitivity, and 96.46\% specificity. A patch-based technique utilizing statistical analysis with ResNet18 and fully convolutional-DenseNet103 architectures is presented in \cite{oh2020deep}. Due to the lack of CXR images for COVID-19, this technique is used with a small dataset to classify COVID-19 patients. Moreover, the proposed technique has achieved an accuracy of 88.9\%  and a precision of 83.4\%. In another study \cite{das2020truncated}, a deep CNNs model is introduced to detect positive patients of COVID-19, called Truncated InceptionNet. The inceptionNetV3 is modified with the Truncated model to deal with limited datasets of COVID-19 to reduce over-fitting. Besides, 162 COVID-19 confirmed cases are added to six several datasets.
The proposed model is evaluated with different datasets to test the ability of the Truncated InceptionNet model for predicting positive cases of COVID-19 of each dataset. This model achieved 99.9\% accuracy for the fifth dataset with 162 COVID-19,1583 normal, and 4280 pneumonia.
The DarkNet technique is presented for COVID-19 classification on CXR images in \cite{ozturk2020automated}. This model is predicted COVID-19 cases without using a feature extractor algorithms. The DarkNet model is applied as a classifier for YOLO object detection. This model is used with binary and multi-class COVID-19 classification. The DarkNet model is achieved 98.08\%, 95.13\%,95.30\%, 96.51\% for accuracy, sensitivity, specificity, and F1-measure, respectively. Besides, the model is achieved 87.02\% of accuracy for a multi-class dataset. In another work \cite{islam2020combined}, a hybrid CNN-LSTM model is presented using CNNs and long short-term memory to detect COVID-19 cases. The CNNs are utilized for feature extraction and LSTM for the classification of images. The CNN-LSTM model is used with 4,575 CXR images, including 1525 COVID-19 images. The CNN-LSTM model has achieved higher results with 99.4\%, 99.3\%, 99.2\%,99.9\%, and 99.9\% for accuracy, sensitivity, specificity, AUC score, and F1-measure, respectively. Also, The AUC-ROC curve and confusion matrix are provided to evaluate the CNN-LSTM model.

In \cite{khan2020coronet}, a deep CNN based on Xception model is introduced, called CoroNet. The proposed model utilized an ImageNet dataset for the pre-trained process. This model is evaluated on binary and multi-class datasets with 284 COVID-19 images. The proposed CoroNet model accomplished an accuracy of 99\%, 95\%,89.6\% for binary, three-class, and four-classes datasets. The overall accuracy is achieved with 89.6\%. In this study \cite{brunese2020explainable}, The VGG16 architecture is used with transfer learning for COVID-19 classification from CXR images. The two models are proposed to classify between (COVID-19 vs. healthy) and (COVID-19 vs. pulmonary) for the first and second models, respectively. The first model achieved 96\% of accuracy and sensitivity, 98\% of specificity, and 94\% of F1-measure. The results 98\%, 87\%, 94\%, and 89\% for accuracy, sensitivity, specificity, and F1-measure, respectively, are achieved by the second model. On the other side, a hybrid GSA-DenseNet121 DL model is presented using a gravitational search optimizer \cite{ezzat2020gsa}. The GSA is used to the hyper$-$parameters tuning of the DenseNet121 model. The random copying is performed to a balance the used dataset. The GSA-DesneNet121 model is applied with 306 CXR images, including 99 positive COVID-19 and 207 Non-COVID-19. The proposed model achieved 98.3\% of accuracy. In a similar work \cite{sahlol2020covid}, a hybrid CNN's technique is proposed using a swarm-based optimizer for COVID-19 classification. The hybrid technique is combined among inception model, Marine Predators, and fractional-order calculus algorithms. The Inception CNNs model is used for feature extraction. The Marine Predators algorithm is applied to reduce extractor features by the selection of meaningful features. The proposed technique (IFM) is improved using the fractional-order calculus algorithm. This technique is evaluated on two COVID-19 datasets: the first dataset includes 200 and the second dataset contains 219  COVID-19 CXR images.  The proposed IFM technique achieved 98.7\% of accuracy and 98.2\% of f1-score for the first dataset. The results 99.6\%of accuracy, and 99\% of f1-score are obtained for the second dataset.

A new approach \cite{toraman2020convolutional} is proposed utilizing capsule network and CXR images for COVID-19 cases detection called the CapsNet model. This model is evaluated on binary and multi-class datasets. This model has used 1050 CXR images for COVID-19, normal, and pneumonia classes. For binary dataset, The CapsNet model is accomplished  97.24\%, 97.42\%, 97.04\%, 97.08\%, and 97.24\% for accuracy, recall, specificity,	precision, and F1-measure, respectively. Also, the results 84.22\% of accuracy, 84.22\% of recall, 91.79\% of specificity, 84.61\% of precision, and 84.21\% of F1-measure are achieved for the multi-class dataset. In another separate study \cite{marques2020automated},  a novel system using the EfficientNet model to diagnose COVID-19 patients. The EfficientNet has used a 10-fold cross-validation technique for binary and multi-class classification with 1,508 a total number of CXR images. This model has reported binary class results  99.62\% of accuracy, 99.63\% recall,  99.64\% of precision, and 99.62\% of f1-score. The model achieved 96.70\%,  96.69\%,  97.59\%, and 97.11\% for accuracy,  recall,  precision, and  F1-score, respectively with multi-class classification.

In \cite{xu2020deep}, a 3D CNNs is introduced based on the location-attention model to distinguish COVID-19 patients from CT images. The dataset images are composed of 219 COVID-19, 224 pneumonia, and 175 normal images. This model is achieved 86.7\% of classification accuracy. In another study with CT images \cite{singh2020classification}, the CNNs model is proposed with multi-objective differential evolution for tuning hyper-parameters of CNNs. The results are reported with 0.92\% accuracy, 0.90\% for sensitivity, specificity, and F1-measure.

The ResNet50 model is used with multi-view images for the classification of COVID-19 patients with chest CT scans. The proposed model is achieved an AUC of 81.9\%, the sensitivity of 81.8\%, accuracy of 76\%, and specificity of 61.5\% in \cite{wu2020deep}. In \cite{jaiswal2020classification}, a deep learning technique is proposed using DenseNet$-$201 model and transfer learning. The proposed technique is used on 2492 CT images, including 1262 COVID-19, and 1230 Non-COVID-19 images. The proposed technique is applied with learned weights of the ImageNet dataset to extract features. It is achieved accuracy with 97\% compared to Resnet152V2, VGG16 models. Besides, it obtained AUC- ROC score with 0.97\%.

The previous studies faced inconsistency issues due to a variety of methods, images, a variety of data. A brief comparison of the previous research has been presented in terms of the applied methods, images, data, and classification performance (as shown in Table \ref{Tbl:T1}). Since chest CXR and CT scans are the second tools to diagnose and detect the COVID-19 pandemic, so the classification of COVID-19 has become a hot area of research. The researchers collected datasets from several resources to train and test the models as shown in Table \ref{Tbl:T1}. There is no a standard dataset due to the shortage of COVID-19 images. Most of the state-of-the-art COVID-19 classification is designed to fit the currently limited images as \cite{oh2020deep,das2020truncated} and to help in early diagnosis for COVID-19 cases. There studies focus on prediction confirmed COVID-19 from COVID-19 and normal datasets such as \cite{ozturk2020automated} and another focus on differentiating confirmed COVID-19 from other pneumonia diseases \cite{das2020truncated,wu2020deep,ezzat2020gsa}. Therefore, the prediction of COVID-19 still needs to further accurate and faster models to assist the radiologists in coming early diagnosis and treatment.

\begin{table*}[!htb]
	\centering
	\caption{ Comparison of classification performance for existing related work.}
	\label{Tbl:T1}\resizebox{1\textwidth}{!}{%
		\begin{tabular}{m{1.5cm} m{7cm} m{1cm} m{4cm}  m{2cm} m{2cm}} \hline
			\textbf{Work} & \textbf{Images}  & \textbf{Type} &\textbf{Method}  &\textbf{Data}& \textbf{ACC.(\%)}\\\hline
			\cite{apostolopoulos2020covid}& 224 COVID-19  714 pneumonia    504 normal &  CXR & MobileNetV2  & imbalanced & 96.78\\ \\
			\cite{oh2020deep} &  180 COVID-19  191 normal 131 others   & CXR &ResNet18 & imbalanced & 88.9 \\ \\
			\cite{das2020truncated} & 162 COVID-19 & CXR & CNNs& imbalanced& 99.9 \\ \\
			\cite{ozturk2020automated} & 125 COVID-19 500 normal& CXR &  DarkNet& imbalanced	& 98.08\\ \\
			\cite{islam2020combined} & 1525  COVID-19   1525 normal  1525 pneumonia& CXR & CNN$+$LSTM   & balanced & 99.4 \\ \\
			\cite{khan2020coronet} & 284 COVID-19    330 bacterial  310 normal  327 Viral  & CXR   & Xception & balanced	& 89.6\\ \\
			\cite{brunese2020explainable} & 250 COVID-19
			2753 pneumonia 
			3520  normal & CXR
			&  VGG16  	& imbalanced	& 96\\ \\
			\cite{ezzat2020gsa} &99 COVID-19 207 Non-COVID-19 &  CXR  & DenseNet121$+$GSA &  balanced& 98.3  \\ \\
			\cite{sahlol2020covid} & {200 COVID-19 1,675 Non-COVID-19} &{ CXR} & {Inception+FO‑MPA}& {imbalanced}&{ 98.7} \\\\
			{\cite{sahlol2020covid}} & {219 COVID-19 1,341 Non-COVID-19} & {CXR} & {Inception+FO‑MPA} & {imbalanced} & {99.6} \\\\
			{\cite{toraman2020convolutional}}& {1050 COVID-19 1050 normal} &{ CXR}& {Capsule Network}& {balanced} & {97.24} \\ \\
			{\cite{toraman2020convolutional}} & {1050 COVID-19 1050 normal 1050 pneumonia} & {CXR}& {Capsule Network}& {balanced} & {84.22} \\ \\
			{\cite{marques2020automated}}& {504 COVID-19 500 normal} &{ CXR}& {EfficientNet}& {balanced} & {99.62} \\ \\
			{\cite{marques2020automated}} & {504 COVID-19 500 normal 504 pneumonia} & {CXR}& {EfficientNet}& {balanced} & {96.70} \\ \\
			\cite{xu2020deep} & 219 COVID-19   175 normal  \hspace{1mm}  224 pneumonia  & CT& 3D-DL & balanced & 86.7 \\ \\
			\cite{singh2020classification} & 73 COVID-19 & CT&  CNNs$+$MODE &  - & 92\\ \\
			\cite{wu2020deep} & 368 COVID-19  127 pneumonia &CT& ResNet50 & imbalanced& 76 \\ \\
			\cite{jaiswal2020classification} & 1262 COVID-19 1230 Non$-$COVID-19 & CT & DenseNet201 & balanced& 97\\ 
			\hline
	\end{tabular}}
\end{table*}

\section{Preliminaries}
\label{Sec:Preliminaries}
In this section, an introduction about some of the concepts quantum computing used in this work like the quantum bit, quantum gates, and architecture of convolutional neural networks (CNNs). If the reader is familiar with these concepts may skip this subsection.

\subsection{Quantum computing}
Quantum computing relies on postulates and characteristics of quantum mechanics (i.e., quantum bits, interference, superposition, and entanglement) to information processing. Quantum computing gives us the ability to solve complex problems better and faster than classical computing \cite{dunjko2016quantum,aimeur2006machine}.
Qubit is the small unit to process information in a quantum computer like the bit in classical computing. A qubit can also be in one-state, zero-state, or both of states at the same time known as linear superposition. The qubit is a state vector in Hilbert space.
\begin{equation}
\label{Eq:eq1}
|\psi\rangle=\left(\begin{array}{l}
\theta \\ 
\delta
\end{array}\right)=\theta|0\rangle+\delta|1\rangle
\end{equation}
Where $\theta$ and $\delta$ are the probability amplitudes that represented by complex numbers and $| \theta^{2} |   + | \delta^{2} | $  = 1

From the postulates of quantum mechanics, any unitary transformation (unitary matrix) is a quantum gate. Matrix to be unitary, the following condition must be verified:
\begin{equation}
U U^{\dagger}=U^{\dagger} U=I 
\end{equation}
where $U^{\dagger}$ is the conjugate transpose of a matrix $U$ and $I$ is an identity matrix. 
Quantum gates can be classified based on numbers of qubits: one-qubit gates, two-qubits gates, and multiple-qubits gates \cite{kaye2007introduction,nielsen2002quantum}. Firstly, from the most popular and widely used gate in one-qubit gates is a Hadamard gate or square-root of NOT gate. The use of the Hadamard gate to present qubits into superposition. Pauli gates also are one-qubit gates. Secondly, two-qubits gates which work on $4 x 4$ unitary matrices. For example, Controlled NOT, and swap gate. Lastly, multiple-qubits gates which work on multiple qubits as $2^{n}$ x $2^{n}$ unitary matrices such as Toffoli and SWAP gates. Table \ref{Tbl:T2} shows quantum gates with circuit representation, unitary matrix, the number of qubits, and the operation of the gate.

\begin{table*}
	\centering
	\caption{summarizes standard quantum computing gates.}
	\label{Tbl:T2} \resizebox{0.92\textwidth}{!}{%
		\begin{tabular}{|m{2cm}|m{4cm}|m{7cm}|m{1cm}|m{4cm}|@{}p{0pt}@{}}\hline
			Gate   & Notation   & Matrix & Qbit & Use& \\ \hline
			%%%%%%%%%%%%% SecondRow_Hadmard gate%%%%%%%%%%
			Hadamard &
			\includegraphics[width=2cm,height=1cm]{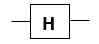}
			&      $$\frac{1}{\sqrt{2}}\left(\begin{array}{cc}
			1 & 1 \\
			1 & -1
			\end{array}\right)$$    & 1 &   Create superposition state between two quantum bits.  \\ \hline
			%%%%%%%%%%%%% 3 Row_Pauli X gate %%%%%%%%%%
			Pauli-X &
			\includegraphics[width=2cm,height=1cm]{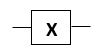} 
			& $$\left(\begin{array}{ll}
			0 & 1 \\
			1 & 0
			\end{array}\right)$$ & 1 &   Flip quantum bit from state to another.  \\ \hline
			%%%%%%%%%%%%% 3 Row_Pauli Y gate %%%%%%%%%%
			Pauli-Y &
			\includegraphics[width=2cm,height=1cm]{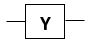} 
			&  $$\left(\begin{array}{cc}
			0 & -i \\
			i & 0
			\end{array}\right)$$    &  1  &   Make $\pi$-rotation for quantum bit around the Y-axis.   \\ \hline
			%%%%%%%%%%%%% 3 Row_Pauli Y gate %%%%%%%%%%
			Pauli-Z &
			\includegraphics[width=2cm,height=1cm]{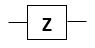} 
			\label{Fig: F0-4}
			&      $$\left(\begin{array}{cc}
			1 & 0 \\
			0 & -1
			\end{array}\right)$$    & 1 & Make  $\pi$ -rotation for quantum bit around the Z-axis.\\ \hline
			%%%%%%%%%%%%% SWAP gate %%%%%%%%%%
			Swap & \includegraphics[width=2cm,height=1.7cm]{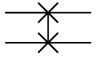} 
			&  $$\left(\begin{array}{llll}
			1 & 0 & 0 & 0 \\
			0 & 0 & 1 & 0 \\
			0 & 1 & 0 & 0 \\
			0 & 0 & 0 & 1
			\end{array}\right)$$  & 2 &  Swap two quantum bits states.  \\ \hline
			%%%%%%%%%%%%% 3 Row_Pauli Y gate %%%%%%%%%%
			Toffoli  &  \includegraphics[width=2.5cm,height=2cm]{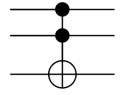}
			&  $$\left(\begin{array}{llllllll}
			1 & 0 & 0 & 0 & 0 & 0 & 0 & 0 \\
			0 & 1 & 0 & 0 & 0 & 0 & 0 & 0 \\
			0 & 0 & 1 & 0 & 0 & 0 & 0 & 0 \\
			0 & 0 & 0 & 1 & 0 & 0 & 0 & 0 \\
			0 & 0 & 0 & 0 & 1 & 0 & 0 & 0 \\
			0 & 0 & 0 & 0 & 0 & 1 & 0 & 0 \\
			0 & 0 & 0 & 0 & 0 & 0 & 0 & 1 \\
			0 & 0 & 0 & 0 & 0 & 0 & 1 & 0
			\end{array}\right)$$   & 3 &   Controlled-Controlled  NOT (CC Not) gate. Flip target quantum bit if both control  two quantum bits equal one.\\ \hline
		\end{tabular}
	}
\end{table*}

\subsection{Convolutional Neural Networks}
\begin{figure*}[!htb]
	\centering
	\includegraphics[scale=0.9]{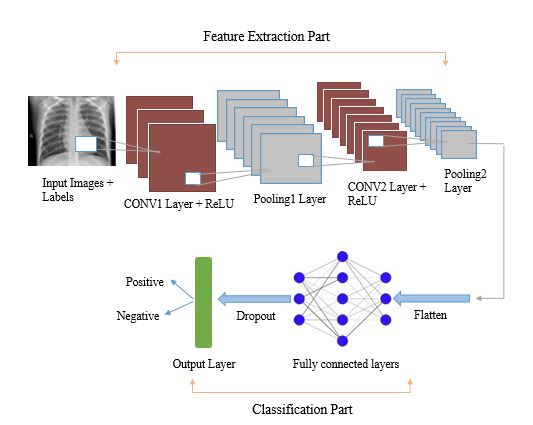}
	\caption{Illustrates the general architecture of the CNNs: The CNNs consists of feature extraction and classification parts. First of all, the input image is pushed to the Conv layer as a matrix of pixels to extract features and patterns from images by apply kernels on the input image. Then, the pooling layer is applied to reduce the number of features and fast training time. The next layer is applied convolution operation on convoluted features with the kernels to extract more invariant and local features and so on. With the last pooled feature map flatten layer has been applied. Finally, fully connected layers have been used for train and classification.}
	\label{Fig:F1}
\end{figure*}

Convolutional Neural Networks (CNNs) \cite{lecun1989backpropagation} inspired by convolution operation to produce a convolutional (Conv) layer that is the heart of the CNNs. As illustrated in Figure  \ref{Fig:F1}, the basic architecture of CNNs is similar to a multi-layer perceptron starts with the input image and sequence of hidden layers to predict labels over the output layer. The layers of CNNs are explained in detail in Section \ref{Sec:Proposed}. The CNNs play a critical role in computer vision applications such as image classification \cite{das2020truncated}, image segmentation \cite{liu2019recent}, object detection \cite{li2019clu}, and signal processing \cite{vrysis20201d}. Various models had been developed based on the concepts of CNNs as ResNet50, and VGG-19.
%%%%%%%%%%%%%%%%%%%%%%%%%%%%%%%%%%%%%%%%%%%%%

\section{Proposed HQCNN Model}
\label{Sec:Proposed}

\begin{figure*}[]
	\centering
	\includegraphics
	[scale=0.8]{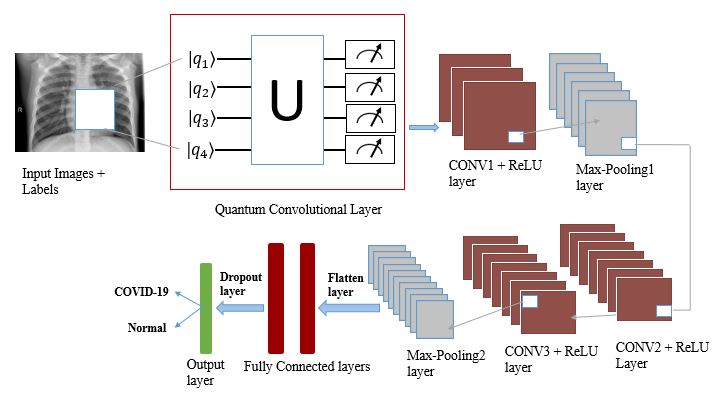} 
	\caption{Block diagram for proposed HQCNN model.}
	\label{Fig:F2}
\end{figure*}

The proposed hybrid quantum-classical CNNs model (HQCNN) aims to improve CNNs classification for medical images and predict COVID-19 and healthy patients in early stages. The main idea of the HQCNN model is based on hybrid computation to enhance the performance of classical learning \cite{bergholm2018pennylane}. The proposed model consists of two parts: first, the quantum part has utilized the quantum Conv layer, which is proposed by Henderson et al. \cite{henderson2020quanvolutional}. Second, the classical part with CNNs structure. As shown in Figure \ref{Fig:F2}, the HQCNN model has one quantum Conv layer, three Conv layers followed by the rectified linear unit activation function, two max-pooling layers, and followed by two fully connected layers. 
The layers of the model are presented in detail in the following steps.

\subsection{Quantum convolutional layer}
The proposed HQCNN model has used quantum convolution that is presented by the Maxwell Henderson et al. \cite{henderson2020quanvolutional}.
Quantum convolution works as small random quantum circuits (RQCs) to calculate convolution operation and can implement on near-term quantum hardware. This circuit has applied with local locations of input images to extract elementary and informative features. The quantum convolution layer consists of three phases: encoding, random quantum circuit, and decoding(as shown in Figure \ref{Fig:F2}).

\subsubsection{Encoding}	
Up till now, the encoding data to quantum is a challenge in quantum machine learning (QML) \cite{abohashima2020classification}. Several encoding methods have been discussed in \cite{lloyd2020quantum}. In this work, angle encoding has been used to transform input data into rotation angles of quantum states. The angle rotations gates are the simplest encoding methods to access data into a quantum circuit. In this study, due to needing the single quantum gate for each entry to encode data. These gates are corresponding to encode classical pixel data to a quantum state. Rotation matrices are rotation operators of Pauli-matrices in the form of exponential around $X$, $Y$, $Z$ axes (as seen in Table \ref{Tbl:T1}). Three rotation gates $Rx$, $Ry$, $Rz$ are a single quantum bit rotation via angle $\alpha$ around $X$, $Y$, and $Z$ axes, respectively. The RY and RX rotation gates can be represented by the following equations.
%%%%%%%%%%%%%% Rx gate
\begin{equation}\label{Rx_gate}
R_{x}(\alpha)=\left(\begin{array}{cc}\cos \left(\frac{\alpha}{2}\right) & -i \sin \left(\frac{\alpha}{2}\right) \\ -i \sin \left(\frac{\alpha}{2}\right) & \cos \left(\frac{\alpha}{2}\right)\end{array}\right)
\end{equation}
%%%%%%%%%%%%%% Ry gate  %%%%%%%%%%%%
\begin{equation}
R_{y}(\alpha)=\left(\begin{array}{cc}\cos \left(\frac{\alpha}{2}\right) & -\sin \left(\frac{\alpha}{2}\right) \\ \sin \left(\frac{\alpha}{2}\right) & \cos \left(\frac{\alpha}{2}\right)\end{array}\right) 
\end{equation}

\subsubsection{Random Quantum Circuit}
A quantum circuit is a series of quantum unitary operations (or gates) and measurements connected via wires (Qubits).
Just like the classical Conv layer, quantum Conv layer composed of quantum kernels apply to the input image. The main idea of quantum convolution utilizing random quantum circuits to split input image into small local locations to extract meaningful features. The advantages of the quantum circuit in quantum convolution works with a few quantum bits and shallow-depth of quantum circuits.

\subsubsection{Measurement}
The measurement phase is also known as the decoding phase. Decoding is measuring quantum data to transform into classical form. Pauli matrices can use as measurement methods \cite{takeuchi2019quantum}. In the HQCNN model, the Pauli-Z gate is used to the decoding phase (as seen in Table \ref{Tbl:T2}).

\subsection{Convolutional layer}
Conv layer is the pivotal and significant layer in the feature extraction part of CNN. This layer performs convolution operation on input features with kernels to extract invariant and informative features from images as convoluted features map to the next layer. Convolution operation computes dot products between small local locations of the input image and kernels\cite{ozturk2020automated,zhang2016deep}.
\begin{equation}
%F^{H}=\sum_{i=1}^{q} W_{i} * x^{t}+b
(F * K)(i, j)=\sum_{m} \sum_{n} K(m, n) F(i-m, j-n)
%c_{i j}=\sum_{k} a_{i k} b_{k j}
\end{equation}
Where $*$ is the convolution operation to produce a convoluted feature map, $F$ is the input feature, and $K$ is kernel or filter. Conv layer has been followed by the ReLU function transformation to add activation values into the model network. The function returns $0$ for all negative values and returns the maximum value for all positive values. The ReLU function can be defined by the following formula:
\begin{equation}
ReLU(x) =max(0,x)  
\end{equation}
Where $x$ is an input value.

\subsection{Max-pooling layer}
Max-pooling layer has used to reduce computational learning by selection the most important and valuable features. Convoluted feature maps have been divided into small regions based on the stride number. The max-pooling idea has taken the maximum value of each small location to produce pooled feature maps. Before the classification part, flatten is used to link two parts in CNNs by converting the max-pooled feature map into a 1-dimensional array.

\subsection{Fully Connected Layer}
The Fully Connected (FC) layer is the second part of the CNNs structure. The fully connected layers perform the classification process after the flattening layer by applied weights to predict classes. The dropout layer is used after the FC layer to reduce the overfitting of the model (as shown in Figure \ref{Fig:F2}). 

\begin{figure*}[]
	\centering
	\includegraphics
	[scale=0.6]{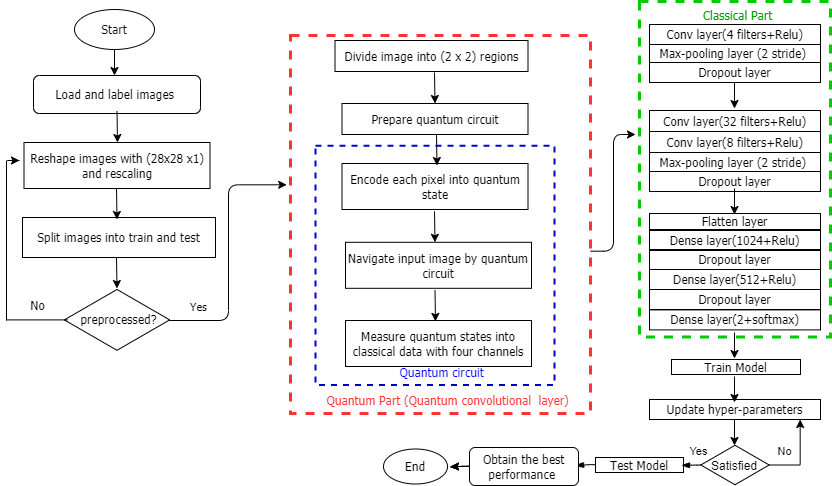} 
	\caption{Flowchart for proposed HQCNN model.}
	\label{Fig:F3}
\end{figure*}

Figure \ref{Fig:F3} illustrates the flowchart of the HQCNN model: the proposed HQCNN model has nine layers, one quantum Conv layer, three classical Conv layers with ReLU function, two max-pooling layers, two FC layers with ReLU function, and one output layer is applied with a softmax activation function. The quantum Conv layer is used encoding method with RY rotation gate and decoding with the Pauli-Z gate. The three classical Conv layers are used with 2D CNNs and are combined with the max-pooling layer. Each max-pooling layer is performed with a 2 x 2 kernel size and the stride equal to $2$, followed by a dropout layer with a 0.2 dropout rate. The two FC layers are applied with 300 and 100 neurons for the first and second layers, respectively. The total parameters for the HQCNN model are 120,394. Table \ref{Tbl:T3} shows the summary layers of the proposed HQCNN model.

\begin{table*}[!htb]
	\centering
	\caption{Shows the summary of the HQCNN model.}
	\label{Tbl:T3}\resizebox{0.8\textwidth}{!}{%
		\begin{tabular}{m{1cm} m{3cm} m{2cm} m{2cm} m{2cm}  m{3cm} m{4cm}} 
			\hline
			\textbf{Layer}	& \textbf{Type}  &\textbf{Units} &\textbf{Kernel Size}& \textbf{Input size}& \textbf{No. of parameters}\\  \hline
			1 & Quantum Conv & 4 & 2 x 2 &   (28 x 28 x 1) & - \\ \\
			2 &  Conv2D  & 16  & 2 x 2 &   (14 x 14 x 4)  & 272 \\ \\
			3& MaxPooling2D&  - & 2 x 2 &  (14 x 14 x 4) &    -\\ \\
			4& Conv2D &      16 & 2 x 2 &    (7 x 7 x4) &    1040\\ \\
			5& Conv2D & 32 & 2 x 2 &    (7 x 7 x 32) &  2080 \\ \\
			6& MaxPooling2D & - &  2 x 2 &  (7 x 7 x 8)& -\\ \\
			7& FC1 &   300 & - &  (288)&  86,700      \\\\
			8& FC2 &   100 & - & (300)&  30,100     \\\\
			9& Output &   2 & - & (100)&  202     \\
			\hline
	\end{tabular}}
\end{table*}

\section{Experimental Results and Discussion}
\label{Sec:Results}
This section introduces the used images dataset, the performance measures to evaluate the HQCNN model, and the analysis and discussion of experimental results.

\subsection{Dataset}
\label{Subsec:data}
\begin{figure}[]
	\centering
	\includegraphics [scale=0.6]{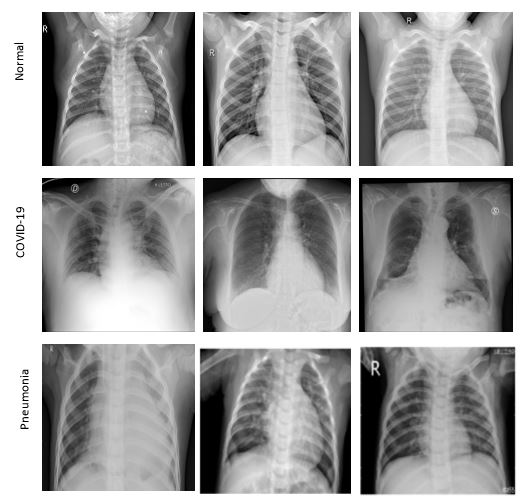} 
	\caption{Samples of used images.}
	\label{fig:F4}
\end{figure}

The used CXR COVID-19 images in this study are collected from two collections. Firstly, COVID-19 image dataset \cite{cohen2020covidProspective} is made by Joseph Cohen. This dataset includes COVID-19, viral, MERS, SARS, and pneumonia. For this study, only 670 CXR COVID-19 images are adapted. Secondly, the augmented COVID-19 dataset\cite{Shafai2020AugmentedCOVID} is made by Walid El-Shafai. This dataset includes CXR  and CT COVID-19 images with augmentation. For this study, only 491 the CXR COVID-19 images without augmentation are adapted. The pneumonia dataset \cite{kermany2018identifying} is used for normal and pneumonia (viral and bacterial) CXR images. The pneumonia dataset includes 5216 train images (1341 normal and 3875 pneumonia) and 624 test images (234 normal and 390 pneumonia). The images are divided into three datasets to evaluate the ability of the proposed HQCNN model to detect COVID-19 cases. The first binary dataset (D1) including 1161 COVID-19 and 1575 normal images. The second binary dataset (D2) includes 1161 COVID-19 and 4247 pneumonia images. The third multi-class dataset (D3) includes 1161 COVID-19, 1575 normal, and 4247 pneumonia images. Table \ref{Tbl:T4} summarizes the used CXR images in this work. Figure \ref{fig:F4} shows samples from the used images.

\begin{table*}[!htb]
	\centering
	\caption{Summarizes of three datasets are used in this work.}
	\label{Tbl:T4}\resizebox{0.8\textwidth}{!}{%
		\begin{tabular}{m{2cm} m{3cm} m{2cm} m{3cm} m{3cm} m{3cm} m{2cm}} 
			\hline
			\textbf{dataset}	& \textbf{COVID-19}  &\textbf{normal} &\textbf{pneumonia}	& \textbf{Train}& \textbf{Test}& \textbf{Total}\\  \hline
			D1   	& 1161	 	& 1575	& -	& 1010 COVID-19   1341 normal    &	151 COVID-19  234 normal & 2736\\ \\
			D2    	& 1161		& -	& 5216 &  1000 COVID-19  3875 normal  & 161 COVID-19 390 pneumonia & 5377\\ \\
			D3      & 1161      &  1575 & 5216 &  1341 normal  1000 COVID-19  3875 pneumonia &     161 COVID-19    390 pneumonia 234 normal   & 6952 \\
			\hline
	\end{tabular}}
\end{table*}

\subsection{Performance measures}
\label{Subsec:perf-measures}
The most common measures for evaluating classification models are confusion matrix with measures metrics like accuracy,  specificity (or true-negative rate), sensitivity (Recall or true-positive rate), precision, and F1-measure. Measures metrics can compute by terms TP, TN, FP, FN as the following formulas:
\begin{itemize}
	\item  True Positives (TP): The proposed HQCNN model correctly predicts COVID-19 cases and labelled as COVID-19.
	\item True Negatives (TN): The proposed HQCNN model correctly predicts normal cases and labelled as normal.  
	\item False Positive (FP): The proposed HQCNN model incorrectly predicts normal cases and labelled as normal.
	\item False Negatives (FN): The proposed HQCNN model incorrectly predicts COVID-19 cases and labelled as COVID-19.
\end{itemize}
\begin{equation}
\label{Eq:acc}
Accuracy (Acc.)= TP+TN / (TP+TN+FP+FN)
\end{equation}
\begin{equation}
\label{Eq:Sens}
Sensitivity (Sns.)= TP/TP+FN
\end{equation}
\begin{equation}
\label{Eq:Spec}
Specificity (Spc.)= TN/TN+FP
\end{equation}
\begin{equation}
\label{Eq:Pre}
Precision (Prc.)= TP/TP+FP
\end{equation}
\begin{equation}
\label{Eq:F1}
F1-measure = 2*(Precision * Recall) / (Precision+Recall)
\end{equation}
\begin{equation}
\label{Eq:B-Acc}
Balanced Accuracy  = 1/2 (TP/TP+FP + TN/TN+FN)
\end{equation}
\begin{equation}
\label{Eq:Fbeta}
\centering
\begin{split}
FBeta-measure = (1+\beta^2) (Precision * Recall)/(\beta^2 * \\  (Precision+Recall))
\end{split}
\end{equation}
\begin{equation}
\label{Eq:FPR}
False  Positive  Rate = FP/FP+TN 
\end{equation}

Receiver Operating Characteristic (ROC) curve identifies as ROC-Area Under Curve (ROC-AUC) is the method to measure the ability of the classifier to predict the right labels. The ROC is a curve between the true-positive rate (sensitivity) and the false-positive rate (1 $-$ sensitivity). Fbeta-measure is the weighted harmonic average of precision and recall of the HQCNN model.

%%%%%%%%%%%%%%%%%%%%%%%%%%%%%%%%%%%%%%%%%%
\subsection{Discussion}
\label{Subsec:discussion}
\begin{figure*}[]
	\centering
	\includegraphics
	[scale=0.7]{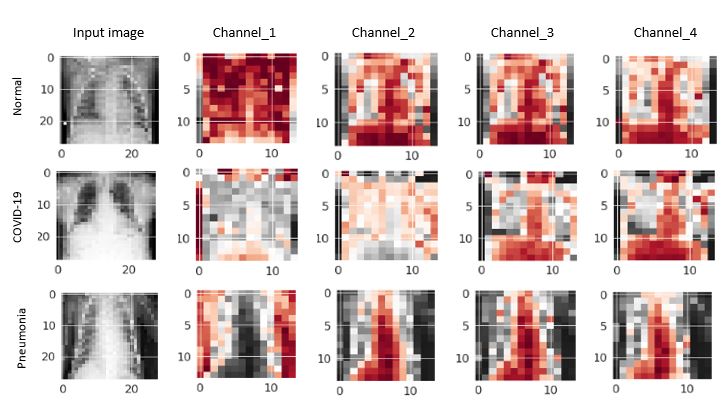} 
	\caption{Samples of normal, COVID-19, and pneumonia images are processed by random quantum circuit.}
	\label{fig:F5}
\end{figure*}

Experimental results have been implemented on Google Colaboratory by using PennyLane and TensorFlow with Keras. PennyLane is a cross-platform Python library for hybrid quantum-classical computation \cite{bergholm2018pennylane}. The main aim of the HQCNN model is to use the concept of the quantum circuit to enhance the performance of the classical CNNs model and the prediction of infected cases of COVID-19 with higher results. RQC works with (2x2) small squares of an input image. Each (2x2) square of the image is encoded into a quantum state by the RY gate. The decoding of the quantum state into classical produces a new four features of a small square. Figure \ref{fig:F5} shows samples of normal, pneumonia, and COVID-19 images, which are examined by the RQC. The first column shows the input image with 28 x 28 size and divided by 250. The quantum Conv layer assigns a feature map into four channels. The quantum Conv layer is used only as a preprocessing layer to CXR images. The train and test phases will be completed on quantum preprocess data with CNNs layers.

At the beginning of the evaluation, the proposed model is evaluated with RY and RX gates to encode data into RQC. RY and RX gates have been used with a different number of shots (500 and 1000). As to be noted from Table \ref{Tbl:T5}, the RY gate (1000 shots) achieved higher results 97.6\%, 99.3\%, 96.5\%, 94.9, and 97\% for accuracy, sensitivity, specificity, precision, and F1-measure, respectively.

\begin{figure}[]
	\centering
	\begin{subfigure}{.35\textwidth}
		\includegraphics[width=1\linewidth]{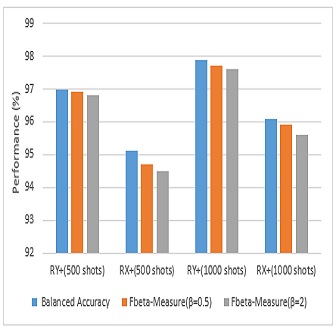}
		\caption{}
		\label{fig:F6-1}
	\end{subfigure}
	\begin{subfigure}{.39\textwidth}
		\centering
		\includegraphics[width=1\linewidth]{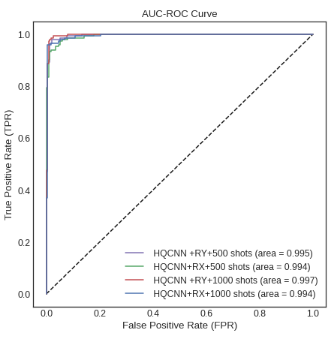}
		\caption{}
		\label{fig:F6-2}
	\end{subfigure}
	\caption{(a) comparative analysis of balanced accuracy, FBeta-measure($\beta$=0.5), and FBeta-measure($\beta$=2) for D1. (b) visualization of the ROC curve.}
	\label{fig:F6}
\end{figure}

\begin{figure}[]
	\centering
	\begin{subfigure}{.4\textwidth}
		\centering
		\includegraphics[width=1\linewidth]{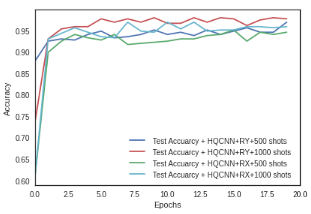}
		\caption{}
		\label{fig:F7-1}
	\end{subfigure}
	\begin{subfigure}{.4\textwidth}
		\centering
		\includegraphics[width=1\linewidth]{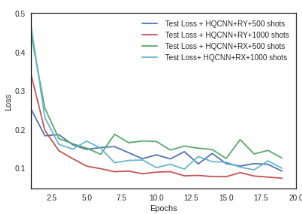}
		\caption{}
		\label{fig:F7-2}
	\end{subfigure}
	\caption{Visualization of the learning curve for both  (a) test accuracy and  (b) test loss for D1 with 20 epochs.}
	\label{fig:F7}
\end{figure}

\begin{figure}[]
	\centering
	\begin{subfigure}{.3\textwidth}
		\centering
		\includegraphics[width=1\linewidth]{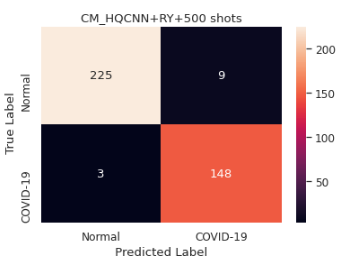} 
		\caption{}
		\label{fig:F8-1}
	\end{subfigure}
	\begin{subfigure}{.3\textwidth}
		\centering
		\includegraphics[width=1\linewidth]{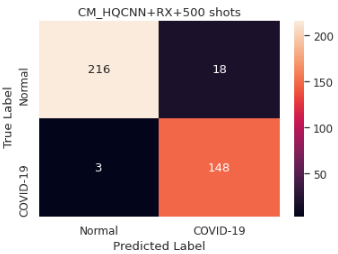} 
		\caption{}
		\label{fig:F8-2}
	\end{subfigure}
	
	\centering
	\begin{subfigure}{.3\textwidth}
		\centering
		\includegraphics[width=1\linewidth]{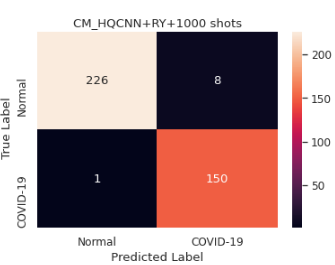} 
		\caption{}
		\label{fig:F8-3}
	\end{subfigure}
	\begin{subfigure}{.3\textwidth}
		\centering
		\includegraphics[width=1\linewidth]{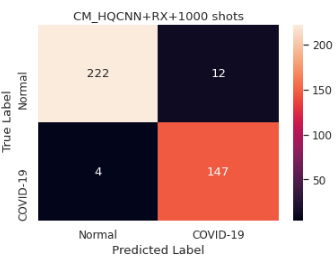} 
		\caption{}
		\label{fig:F8-4}
	\end{subfigure}
	\caption{The confusion matrix of the proposed HQCNN with RY and RX gates. (a) RY(500 shots), (b) RX(500 shots), (c) RY (1000 shots), and (d) RX (1000 shots).}
	\label{fig:F8}
\end{figure}

\begin{table*}[!htb]
	\centering
	\caption{The performance of HQCNN model with rotation gates and number of shots on D1.}
	\label{Tbl:T5}\resizebox{1\textwidth}{!}{%
		\begin{tabular}{ccccccc} \hline
			\textbf{Rotation gate}& \textbf{Shots} & \textbf{ACC.(\%)}	& \textbf{Sns.(\%)}& \textbf{Spc.(\%)}& \textbf{Prc.(\%)}& \textbf{F1-measure(\%)}\\  \hline
			\textbf{RY} &  \multirow{2}{*}{1000} & \textbf{97.6} & \textbf{99.3} & \textbf{96.5}&  \textbf{94.9} & \textbf{97}\\
			RX &            & 95.8 & 97.3 & 94.8& 92.4 & 94.8\\ \hline
			RY &  \multirow{2}{*}{500} & 96.8 & 98 & 96.1 & 94.2& 96.1\\
			RX &                     & 94.5 & 98 & 92.3 & 89.1& 96.1\\
			\hline
		\end{tabular}
	}
\end{table*}
As shown in Figure \ref{fig:F6-1}, the RY gate (1000 shots) achieved 97.9 \% of balanced accuracy. Besides, the RY gate (1000 shots) obtained the largest AUC-ROC with 99.7\% closer to $1$. The test loss curve of the HQCNN model with RY (1000 shots) continues to reduce until the end of the test phase. The test loss with RY (1000 shots) is 0.07\%. In addition, the test loss with RY (500 shots) is 0.09\%, the RX achieved 0.9\% and 0.1\% of test loss for $1000$ and $500$ shots, respectively (see Figure \ref{fig:F7}). Figure \ref{fig:F8} shows the confusion matrix for RY and RX with a different number of shots. The RY (1000 shots) has predicted the highest number of positive cases of COVID-19 (150 of 151 cases), as shown in Figure \ref{fig:F8-3}. Increasing the number of shots will calculate the better estimation of the output statistic of observables (i.e., expectation value and variance).

As mentioned above, increasing the number of shots increases the performance of the model. Besides, the encoding method (RY) can be improved the overall results of the model. So, the experimental results will be completed with an encoding angle (RY gate) and 1000 shots.

The proposed HQCNN model results are compared with the CNNs, multi-layer perceptron (MLP), support vector machines(SVMs), k-nearest neighbours (KNN), AdaBoost, random forest (RF), bagging classifier (BC), decision tree (DT), Gaussian Naive Bayes (GNB), and XGBoost classifiers. In classical CNN, the quantum layer has been replaced by the classical Conv layer with four filters. The HQCNN model is optimized with the Adam algorithm with a learning rate of 0.0001. The performance of HQCNN model is evaluated with measures metrics as represented in Equations \ref{Eq:acc}, \ref{Eq:Sens}, \ref{Eq:Spec}, \ref{Eq:Pre}, and \ref{Eq:F1}. Besides, balanced accuracy \ref{Eq:B-Acc} to calculate the accuracy in an imbalanced dataset, Fbeta-measure \ref{Eq:Fbeta}, and the AUC-ROC curve. The experimental results of HQCNN are divided into four parts to measure the ability of the proposed model for the classification of COVID-19 patients. The first part presents the classification results of the HQCNN model between normal and COVID-19 cases \ref{Subsec:results_D1}. The second part discusses the experimental results of the HQCNN model between COVID-19 and pneumonia patients \ref{Subsec:results_D2}. The last part debates the performance of the HQCNN model for a multi-class dataset among normal, COVID-19, and pneumonia patients \ref{Subsec:results_D3}. The last part compares the HQCNN model to CXR existing studies in this work \ref{Subsec:results_existingstudies}.

\subsubsection{The experimental results on D1} 
\label{Subsec:results_D1}
\begin{figure*}[]
	\centering
	\begin{subfigure}{.3\textwidth}
		\centering
		% include first image
		\includegraphics[width=1\linewidth]{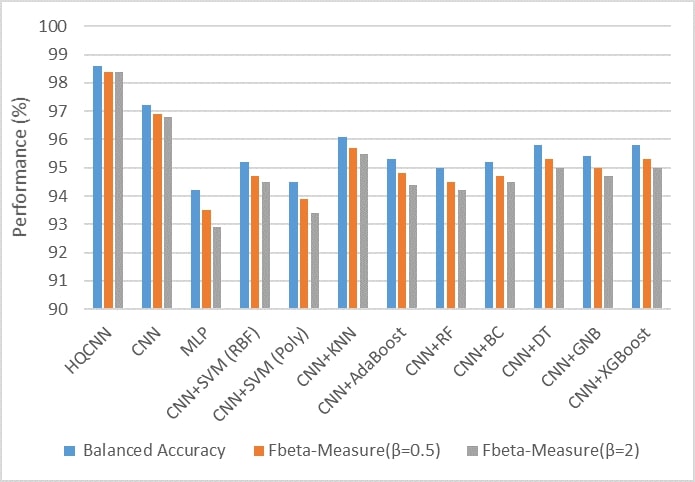} 
		\caption{}
		\label{fig:F9-1}
	\end{subfigure}
	\begin{subfigure}{.3\textwidth}
		\centering
		% include second image
		\includegraphics[width=1\linewidth]{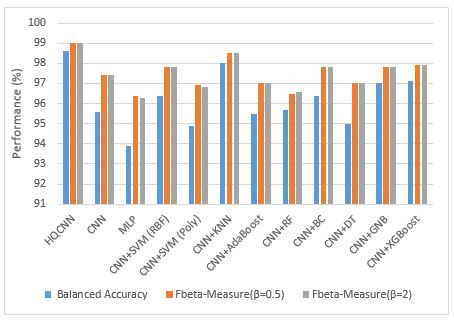}
		\caption{}
		\label{fig:F9-2}
	\end{subfigure}
	\begin{subfigure}{.3\textwidth}
		\centering
		% include second image
		\includegraphics[width=1\linewidth]{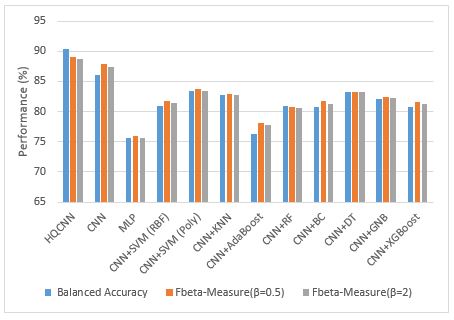}
		\caption{}
		\label{fig:F9-3}
	\end{subfigure}
	\caption{Comparative analysis of balanced accuracy, FBeta-measure($\beta$=0.5), and FBeta-measure($\beta$=2) for (a) D1, (b) D2, and (c) D3.}
	\label{fig:F9}
\end{figure*}

\begin{figure*}[]
	\begin{subfigure}{0.3\textwidth}
		\centering
		\includegraphics[width=.8\linewidth]{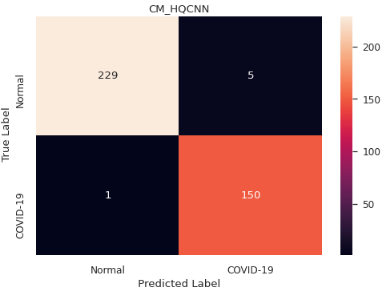}
		\caption{}
		\label{fig:F11-1}
	\end{subfigure}%
	\begin{subfigure}{0.3\textwidth}
		\centering
		\includegraphics[width=.8\linewidth]{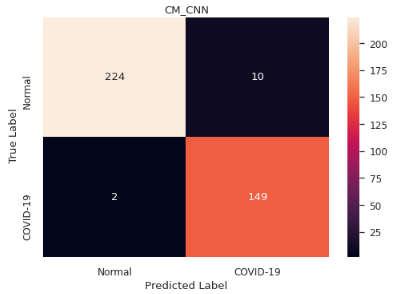}
		\caption{}
		\label{fig:F11-2}
	\end{subfigure}
	\begin{subfigure}{0.3\textwidth}\quad
		\centering
		\includegraphics[width=.8\linewidth]{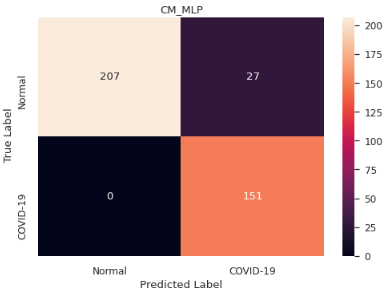}
		\caption{}
		\label{fig:F11-3}
	\end{subfigure}
	\medskip
	
	\begin{subfigure}{0.3\textwidth}
		\centering
		\includegraphics[width=.8\linewidth]{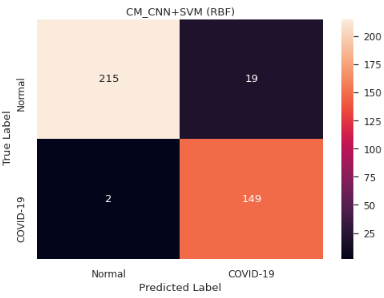}
		\caption{}
		\label{fig:F11-4}
	\end{subfigure}
	\begin{subfigure}{0.3\textwidth}
		\centering
		\includegraphics[width=.8\linewidth]{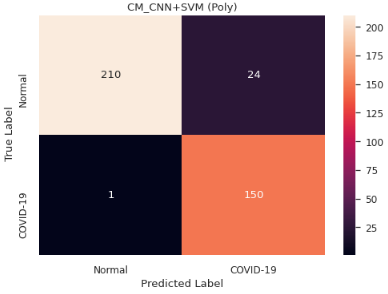}
		\caption{}
		\label{fig:F11-5}
	\end{subfigure}
	\begin{subfigure}{0.3\textwidth}
		\centering
		\includegraphics[width=.8\linewidth]{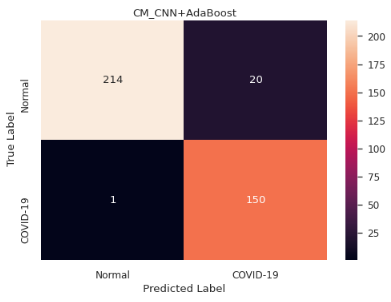}
		\caption{}
		\label{fig:F11-6}
	\end{subfigure}
	
	\begin{subfigure}{0.3\textwidth}
		\centering
		\includegraphics[width=.8\linewidth]{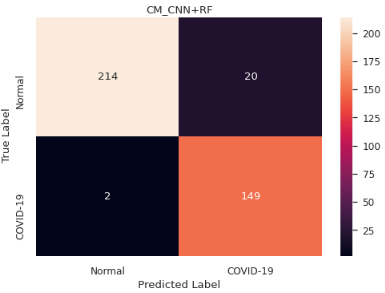}
		\caption{}
		\label{fig:F11-7}
	\end{subfigure}
	\begin{subfigure}{0.3\textwidth}
		\centering
		\includegraphics[width=.8\linewidth]{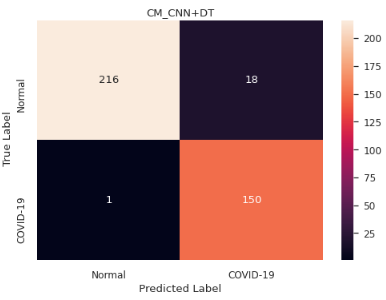}
		\caption{}
		\label{fig:F11-8}
	\end{subfigure}
	\begin{subfigure}{0.3\textwidth}
		\centering
		\includegraphics[width=.8\linewidth]{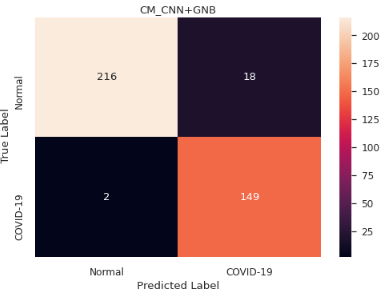}
		\caption{}
		\label{fig:F11-9}
	\end{subfigure}
	
	\begin{subfigure}{0.3\textwidth}
		\centering
		\includegraphics[width=.8\linewidth]{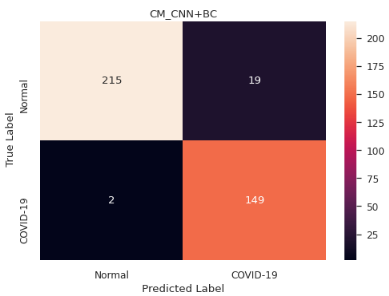}
		\caption{}
		\label{fig:F11-10}
	\end{subfigure}
	\begin{subfigure}{0.3\textwidth}
		\centering
		\includegraphics[width=.8\linewidth]{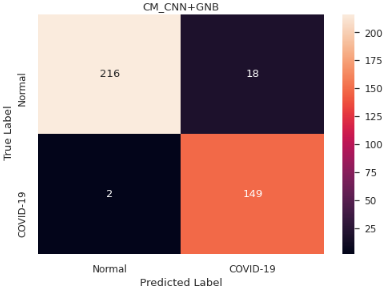}
		\caption{}
		\label{fig:F11-11}
	\end{subfigure}
	\begin{subfigure}{0.3\textwidth}
		\centering
		\includegraphics[width=.8\linewidth]{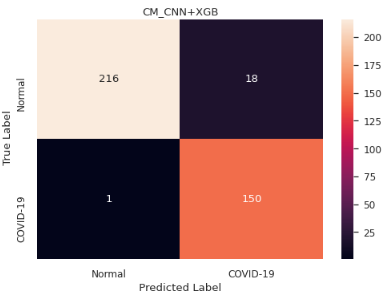}
		\caption{}
		\label{fig:F11-12}
	\end{subfigure}
	\caption{The confusion matrix of the proposed HQCNN and different models for D1.}
	\label{fig:F11}
\end{figure*}

As shown in Table \ref{Tbl:T6}, the performance HQCNN model outperforms CNN and other models. The hybrid proposed model has predicted COVID-19, and normal patients with 98.4\% of accuracy, 99.3\% of sensitivity, 97.8\%  of specificity, 96.7\% of precision, and  98\% of F1-measure. The MLP model outperforms the HQCNN model in measure of sensitivity, it achieved a sensitivity of 100\%. Besides, as observed from Figure \ref{fig:F9-1} proposed hybrid model has achieved higher balanced accuracy and Fbeta-measure with 98.6\% and 98.4\%, respectively. Additionally, The HQCNN model scored the largest AUC of  ROC curve with 0.99\%(near to 1) compared by various models, as shown in Figure \ref{fig:F10-1}. The confusion matrices of binary D1 is displayed in Figure \ref{fig:F11}. The confusion matrix of the HQCNN model, among of 151 COVID-19 images, one patient is misclassified by the HQCNN model. From a total of 234 normal cases, five cases are misclassified by the HQCNN model.

\begin{table*}[!htb]
	\centering
	\caption{Comparison of classification performance between proposed HQCNN and different models on D1.}
	\label{Tbl:T6}\resizebox{.8\textwidth}{!}{%
		\begin{tabular}{cccccc}
			\hline
			\textbf{Model}	& \textbf{ACC.(\%)}	& \textbf{Sns.(\%)}& \textbf{Spc.(\%)}& \textbf{Prc.(\%)}& \textbf{F1-measure(\%)}\\  \hline
			\textbf{HQCNN}  &  \textbf{98.4}	& 99.3&  \textbf{97.8} & \textbf{96.7}&  \textbf {98}\\
			CNN    	 & 96.8   & 98.6	& 95.7   & 93.7  &  96\\ 
			MLP     & 92.9 & \textbf{100}& 88.4 & 84.8 & 91.7    \\   
			CNN+SVM(RBF) & 94.5 & 98.6  &  91.8 & 88.6  &  93.4\\
			CNN+SVM(Poly) & 93.5  & 99.3 & 89.7 &  86.2& 92.3\\
			CNN+KNN & 95.5 &  98.6 &  93.5 &   90.8  & 94.6\\
			CNN+AdaBoost &  94.5 &  99.3  & 91.4   &  88.2 & 93.4\\
			CNN+RF & 94.2  & 98.6 &   91.4  &    88.1  &  93.1 \\
			CNN+BC & 94.5  & 98.6  & 91.8  &  88.6 &   93.4   \\
			CNN+DT  &  95  & 99.3  &92.3 &    89.2  & 94\\
			CNN+GNB &    94.8 &98.6 &   92.3  &  89.2 &  93.7  \\
			CNN+XGBoost   &  95 &  99.3  & 92.3  & 89.2 &   94\\\hline
	\end{tabular}}
\end{table*}

\subsubsection{The experimental results on D2}
\label{Subsec:results_D2}
\begin{figure}[]
	\centering
	\begin{subfigure}{.4\textwidth}
		\centering
		% include first image
		\includegraphics[width=1\linewidth]{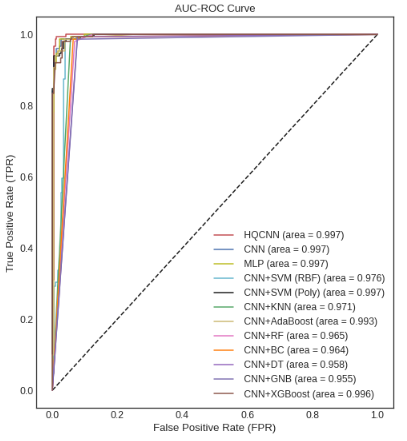} 
		\caption{}
		\label{fig:F10-1}
	\end{subfigure}
	\begin{subfigure}{.38\textwidth}
		\centering
		% include second image
		\includegraphics[width=1\linewidth]{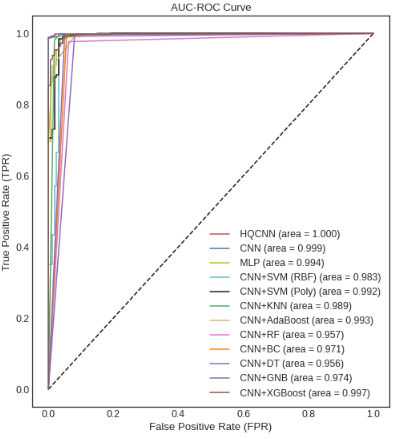} 
		\caption{}
		\label{fig:F10-2}
	\end{subfigure}
	\caption{The AUC-ROC curve of QHCNN and different models on (a) D1, and (b) D2.}
	\label{fig:F10}
\end{figure}

\begin{figure*}[]
	\begin{subfigure}{0.3\textwidth}
		\centering
		\includegraphics[width=.8\linewidth]{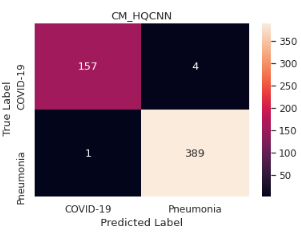}
		\caption{}
		\label{fig:F12-1}
	\end{subfigure}%
	\begin{subfigure}{0.3\textwidth}
		\centering
		\includegraphics[width=.8\linewidth]{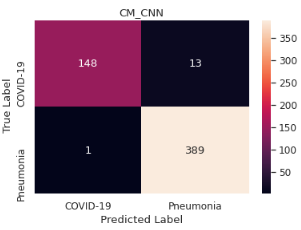}
		\caption{}
		\label{fig:F12-2}
	\end{subfigure}
	\begin{subfigure}{0.3\textwidth}\quad
		\centering
		\includegraphics[width=.8\linewidth]{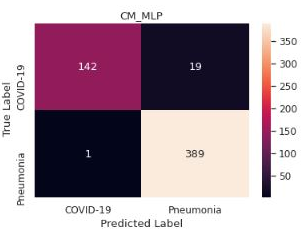}
		\caption{}
		\label{fig:F12-3}
	\end{subfigure}
	\medskip
	
	\begin{subfigure}{0.3\textwidth}
		\centering
		\includegraphics[width=.8\linewidth]{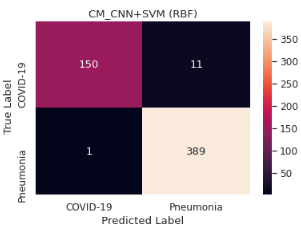}
		\caption{}
		\label{fig:F12-4}
	\end{subfigure}
	\begin{subfigure}{0.3\textwidth}
		\centering
		\includegraphics[width=.8\linewidth]{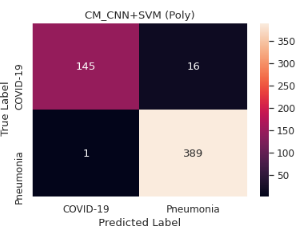}
		\caption{}
		\label{fig:F12-5}
	\end{subfigure}
	\begin{subfigure}{0.3\textwidth}
		\centering
		\includegraphics[width=.8\linewidth]{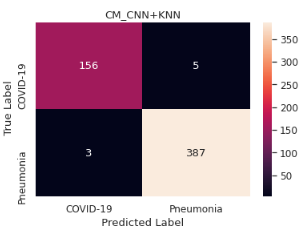}
		\caption{}
		\label{fig:F12-6}
	\end{subfigure}
	
	\begin{subfigure}{0.3\textwidth}
		\centering
		\includegraphics[width=.8\linewidth]{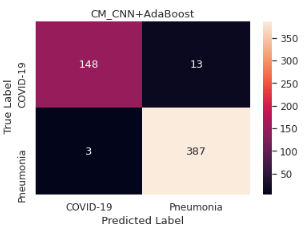}
		\caption{}
		\label{fig:F12-7}
	\end{subfigure}
	\begin{subfigure}{0.3\textwidth}
		\centering
		\includegraphics[width=.8\linewidth]{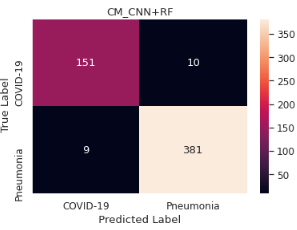}
		\caption{}
		\label{fig:F12-8}
	\end{subfigure}
	\begin{subfigure}{0.3\textwidth}
		\centering
		\includegraphics[width=.8\linewidth]{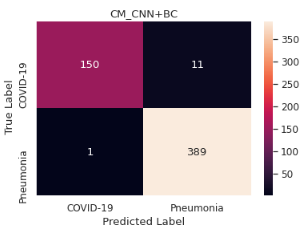}
		\caption{}
		\label{fig:F12-9}
	\end{subfigure}
	
	\begin{subfigure}{0.3\textwidth}
		\centering
		\includegraphics[width=.8\linewidth]{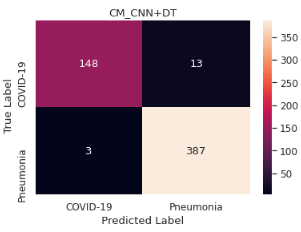}
		\caption{}
		\label{fig:F12-10}
	\end{subfigure}
	\begin{subfigure}{0.3\textwidth}
		\centering
		\includegraphics[width=.8\linewidth]{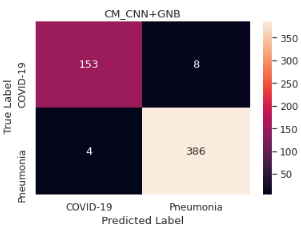}
		\caption{}
		\label{fig:F12-11}
	\end{subfigure}
	\begin{subfigure}{0.3\textwidth}
		\centering
		\includegraphics[width=.8\linewidth]{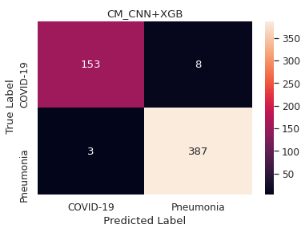}
		\caption{}
		\label{fig:F12-12}
	\end{subfigure}
	\caption{The confusion matrix of the proposed HQCNN and different models on D2.}
	\label{fig:F12}
\end{figure*}

Here, Table \ref{Tbl:T7} reports the results of the HQCNN model compared to various models on D2. the HQCNN model achieved 99\%, 99.3\%, 98.9\%, 99.7\%, and 97.5\% for accuracy, F1-measure, precision, sensitivity, and specificity, respectively. Besides, better-balanced accuracy has been achieved with 98.6\%, 98\% for HQCNN, and CNN+KNN models, respectively as shown in Figure \ref{fig:F9-2}. From Figure \ref{fig:F10-2}, the proposed HQCNN model achieved AUC-ROC score  100\% (equal to 1).
The HQCNN model distinguished between COVID-19 and pneumonia with higher efficiency. The confusion matrices of D2 have been shown in Figure \ref{fig:F12}. From a total of 161 COVID-19 cases, four cases are misclassified by the HQCNN model. Among 390 pneumonia images, one patient is misclassified by the HQCNN model.

\begin{table}[!htb]
	\centering
	\caption{Comparison of classification performance between proposed HQCNN and different models on D2.}
	\label{Tbl:T7}\resizebox{.5\textwidth}{!}{%
	\begin{tabular}{cccccc}\hline
	\textbf{Model}	& \textbf{ACC.(\%)}	& \textbf{Sns.(\%)}& \textbf{Spc.(\%)}& \textbf{Prc.(\%)}& \textbf{F1-measure(\%)}\\  \hline
	\textbf{HQCNN} & \textbf{99}& 99.7&  \textbf{97.5} & \textbf{98.9}& \textbf{99.3}\\
	CNN  &   97.4& 99.7&  91.9 & 96.7  &  98.2\\ 
	MLP  & 96.3 & 99.7& 88.1 & 95.3 & 97.4    \\   
	CNN+SVM(RBF) & 97.8 & 99.7  &  93.1 & 97.2  &  98.4\\
	CNN+SVM(Poly) & 96.9  & 99.7 & 90 & 96 & 97.8\\
	CNN+KNN & 98.5 &  99.2 &  96.8 &  98.7& 98.9\\
	CNN+AdaBoost &  97 &  99.2  &  91.9  &  96.7 & 97.9\\
	CNN+RF & 96.5  & 97.6 &    93.7 &  97.4 &  97.5 \\
	CNN+BC & 97.8  & 99.7  &  93.1 &  97.2 &   98.4   \\
	CNN+DT &  97  & 99.2  & 91.9 &   96.7  & 97.9\\
	CNN+GNB &    97.8 &98.9 &  95  &  97.9 &  98.4 \\
	CNN+XGBoost   &  98  & 99.2  &  95 &   97.9  & 98.5\\ \hline
	\end{tabular}}
\end{table}

%%%%%%%%%

%%%%%%%%%%%%%%%%  Multi-class  Data-set %%%%%%%%%%%%%%%%%
\subsubsection{The experimental results on D3}
\label{Subsec:results_D3}
\begin{figure*}[!h]
	\centering
	\begin{subfigure}{.32\textwidth}
		\centering
		% include first image
		\includegraphics[width=.8\linewidth]{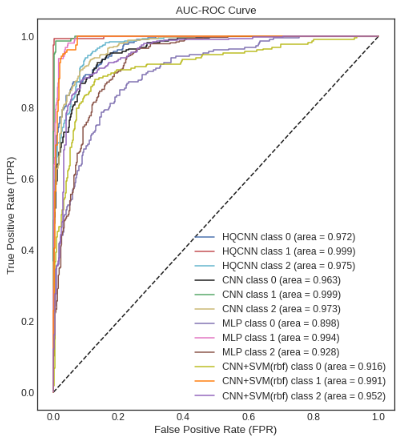} 
		\caption{}
		\label{fig:F13-1}
	\end{subfigure}
	\begin{subfigure}{.32\textwidth}
		\centering
		% include second image
		\includegraphics[width=.8\linewidth]{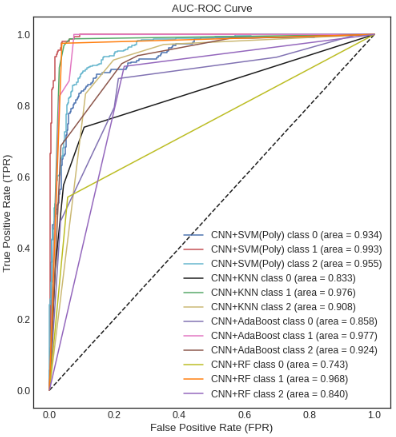}
		\caption{}
		\label{fig:F13-2}
	\end{subfigure}
	\centering
	\begin{subfigure}{.32\textwidth}
		\includegraphics[width=.8\linewidth]{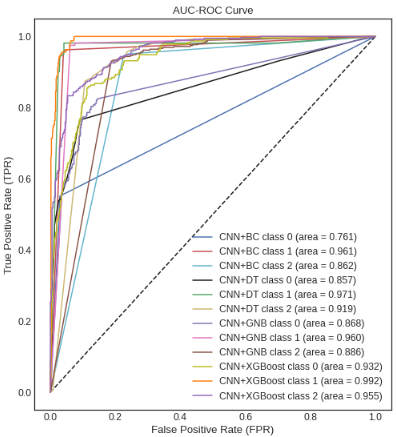}
		\caption{}
		\label{fig:F13-3}
	\end{subfigure}
	\caption{The AUC-ROC curve of QHCNN and different models for D3.}
	\label{fig:F13}
\end{figure*}

\begin{figure*}[!h]
	\centering
	\begin{subfigure}{0.32\textwidth}
		\centering
		\includegraphics[width=.7\linewidth]{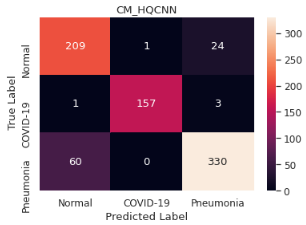}
		\caption{}
		\label{fig:F14-1}
	\end{subfigure}%
	\begin{subfigure}{0.32\textwidth}
		\centering
		\includegraphics[width=.7\linewidth]{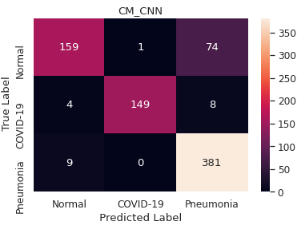}
		\caption{}
		\label{fig:F14-2}
	\end{subfigure}
	\begin{subfigure}{0.32\textwidth}\quad
		\centering
		\includegraphics[width=.7\linewidth]{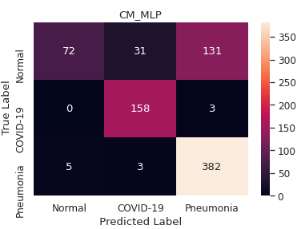}
		\caption{}
		\label{fig:F14-3}
	\end{subfigure}
	\begin{subfigure}{0.32\textwidth}
		\centering
		\includegraphics[width=.8\linewidth]{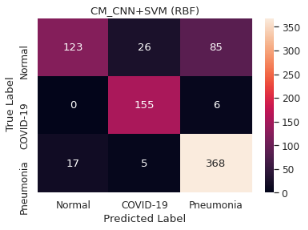}
		\caption{}
		\label{fig:F14-4}
	\end{subfigure}
	\begin{subfigure}{0.32\textwidth}
		\centering
		\includegraphics[width=.8\linewidth]{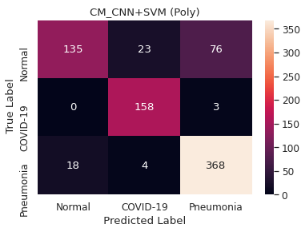}
		\caption{}
		\label{fig:F14-5}
	\end{subfigure}
	\begin{subfigure}{0.32\textwidth}
		\centering
		\includegraphics[width=.8\linewidth]{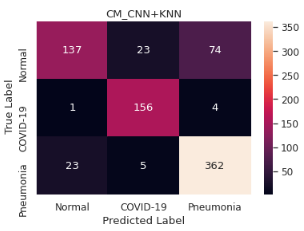}
		\caption{}
		\label{fig:F14-6}
	\end{subfigure}
	\begin{subfigure}{0.32\textwidth}
		\centering
		\includegraphics[width=.8\linewidth]{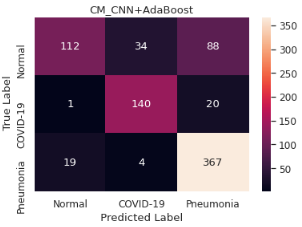}
		\caption{}
		\label{fig:F14-7}
	\end{subfigure}
	\begin{subfigure}{0.32\textwidth}
		\centering
		\includegraphics[width=.8\linewidth]{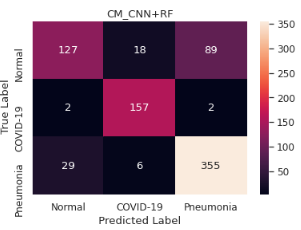}
		\caption{}
		\label{fig:F14-8}
	\end{subfigure}
	\begin{subfigure}{0.32\textwidth}
		\centering
		\includegraphics[width=.8\linewidth]{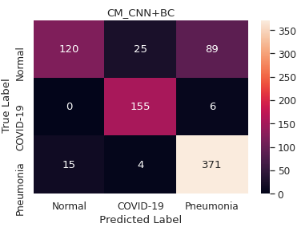}
		\caption{}
		\label{fig:F14-9}
	\end{subfigure}
	\begin{subfigure}{0.32\textwidth}
		\centering
		\includegraphics[width=.8\linewidth]{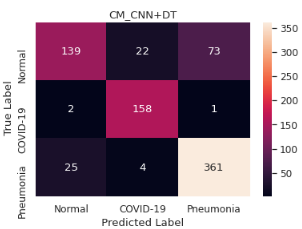}
		\caption{}
		\label{fig:F14-10}
	\end{subfigure}
	\begin{subfigure}{0.32\textwidth}
		\centering
		\includegraphics[width=.8\linewidth]{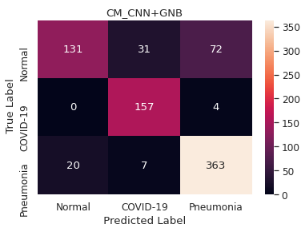}
		\caption{}
		\label{fig:F14-11}
	\end{subfigure}
	\begin{subfigure}{0.32\textwidth}
		\centering
		\includegraphics[width=.8\linewidth]{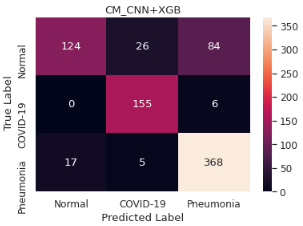}
		\caption{}
		\label{fig:F14-12}
	\end{subfigure}
	\caption{The confusion matrix of the proposed HQCNN and different models on D3.}
	\label{fig:F14}
\end{figure*}

To further evaluate, the imbalanced multi-class dataset is used to evaluate HQCNN performance. The dataset is combined with normal, COVID-19, and pneumonia images. Table \ref{Tbl:T8} presents the classification results of the HQCNN model and other models on an imbalanced multi-class dataset. The best results have been evaluated with 88.6\%, 88.7\%, 89.3\%, and 88.8\% for accuracy, sensitivity, precision, and F1-measure. Besides, the HQCNN model achieved 90.4\% of balanced accuracy compared to various models. In Figure \ref{fig:F13}, The AUC-ROC curve has been exhibited with three normal, COVID-19, and pneumonia classes. It can be noted that the HQCNN distinguished among three classes with higher AUC values 0.97\%, 0.99\%, and 0.97\% for normal, COVID-19, and pneumonia classes. The confusion matrices are shown in Figure \ref{fig:F14}. The confusion matrix of the HQCNN model, among 161 COVID-19 images, four patients are misclassified. From a total of 234 normal cases, 25 cases are misclassified. From 390 pneumonia patients, 60 patients are misclassified by the HQCNN model.

\begin{table}[]
	\centering
	\caption{Comparison of classification performance between proposed HQCNN and different models on D3.}
	\label{Tbl:T8}\resizebox{.5\textwidth}{!}{%
		\begin{tabular}{cccccc}
			\hline
			\textbf{Model}	& \textbf{ACC.(\%)}	& \textbf{Sns.(\%)}& \textbf{Prc.(\%)}& \textbf{F1-measure(\%)}\\  \hline
			\textbf{HQCNN} & \textbf{88.6} & \textbf{88.7} & \textbf{89.3} & \textbf{88.8}\\
			CNN    	 &   87.7& 87.7 & 88.8  &  88.2\\ 
			MLP     &77.9 & 78 & 81.5 & 74   \\   
			CNN+SVM(RBF) & 82.3 & 82.2  & 83.1 &  80.9\\
			CNN+SVM(Poly) & 84.2 & 84.2 &  84.7& 83.2\\
			CNN+KNN & 83.4 &  82 &  83.6  & 82.5\\
			CNN+AdaBoost &  78.8 &  78.8 &  79.8 & 77.3\\
			CNN+RF & 81.4  & 81.2 & 81.2 &  80.3 \\
			CNN+BC & 82 & 82.2  &   83.3 &   80.8   \\
			CNN+DT  &  83.8  & 83  &  83.8  & 82.9\\
			CNN+GNB &   82.9 &82.9 &   83.4 & 81.8 \\
			CNN+XGBoost   &  82.2  & 82.1  & 83.3  & 80.7\\
			\hline
	\end{tabular}}
\end{table}

\subsubsection{Comparison with CXR existing studies}
\label{Subsec:results_existingstudies}
To boost the effectiveness of the HQCNN model in classifying the COVID-19 cases, the HQCNN model is compared to CXR existing studies, as shown in Table \ref{Tbl:T9}. The HQCNN model outperforms previous studies with CXR images in terms of sensitivity and F1-measure. The Truncated InceptionNet and CNN+LSTM models outperform the HQCNN model with a term of accuracy. The Truncated InceptionNet achieved 99.9\% accuracy for one dataset but, the overall accuracy was 98.77\%. The Truncated InceptionNet is based on a large InceptionNetV3 model. The CNN+LSTM has 20 layers, including 12 Conv layers. The HQCNN model achieved better results with nine layers, including four Conv layers (one quantum layer and three classical layers).

\begin{table*}[!h]
	\centering
	\caption{Comparison with CXR existing studies.}
	\label{Tbl:T9}\resizebox{1\textwidth}{!}{%
		\begin{tabular}{m{3.2cm} m{3cm} m{2cm} m{2cm} m{2cm} m{4cm}}	\hline
			\textbf{Study} & \textbf{COVID-19 Images}& \textbf{Class}	& \textbf{ACC.(\%)}	&\textbf{Sns.(\%)}&\textbf{F1-measure(\%)} \\ \hline
			MobileNetV2 \cite{apostolopoulos2020covid} & 224& multi & 96.7& 98.66 & - \\
			ResNet18 \cite{oh2020deep}& 180  & multi&88.9 & 85.9& 84.4 \\ 
			Truncated InceptionNet \cite{das2020truncated} & 162 & multi&99.9 &98 & 98\\
			DarkNet \cite{ozturk2020automated} & 125& binary &98.08 & 95.13 &96.51\\ 
			DarkNet \cite{ozturk2020automated} & 125& multi &87.02 & 85.35& 87.37 \\ 
			CNN+LSTM \cite{islam2020combined}& 1512& binary & 99.4 &99.3 & 99.9 \\
			CoroNet \cite{khan2020coronet} & 280& binary&99& 99.3& 98.5\\ 
			CoroNet \cite{khan2020coronet} & 280& multi&95& 96.9& 95.6 \\
			VGG16 \cite{brunese2020explainable} & 250& binary &96 &96 &94 \\ 
			GSA-DenseNet121 \cite{ezzat2020gsa}&99 & binary &98.38&98.5& 98\\
			{Inception+FO-MPA\cite{sahlol2020covid}} & {200} &{  binary}& {98.7} & {-} &{98.2} \\ 
			{Inception+FO-MPA\cite{sahlol2020covid} }& {219} & {binary} &{99.6} &{-}&{99} \\
			{ CapsNet\cite{toraman2020convolutional}}& {1050}& {binary} &{ 97.24} & {97.42} &{97.24} \\ 
			{CapsNet\cite{toraman2020convolutional}} &{1050} &{ multi} & {84.22}& {84.22}&{84.21} \\ 
			{ EfficientNet\cite{marques2020automated}}& {504}& {binary} &{ 99.62} & {99.63} &{99.62} \\ 
			{EfficientNet\cite{marques2020automated}} &{504} &{ multi} & {96.70}& {96.69}&{97.11} \\ 
			\textbf{HQCNN} & 1161& binary &99 & 99.7& 99.3\\ 
			\textbf{HQCNN} & 1161& multi &88.6 & 88.7& 88.8\\ \hline
		\end{tabular}
	}
\end{table*}

In summary, as mentioned in \ref{Subsec:results_D1}, \ref{Subsec:results_D2}, \ref{Subsec:results_D3}, and \ref{Subsec:results_existingstudies}, the proposed HQCNN model achieved a sensitivity of 99.3\%, 99.7\%, and 88.7\% on D1, D2, D3, respectively. Table \ref{Tbl:T10} shows the results of the HQCNN model with each class on D1, D2, and D3. In D1, the COVID-19 class is predicted with high accuracy of 99.3\%, sensitivity of 97\%, and F1-measure of 98\%. In normal cases, The HQCNN model obtained 100\% sensitivity. In imbalanced binary D2, the COVID-19 class is categorized with 97.5\%, 99\%, and 98\% for accuracy, sensitivity, and F1-measure, respectively. The pneumonia class achieved 100\% precision, 99.7\% accuracy, and 99\% for sensitivity and F1-measure. Furthermore, in the imbalanced multi-class D3, the higher results are obtained with COVID-19 class. It classified the COVID-19 cases with an accuracy of 97.5\%, sensitivity of 99\%, and F1-measure of 98\%. Besides, with pneumonia category achieved a high 92\% sensitivity. The lower sensitivity is achieved with 77\% in normal cases. The higher precision is achieved with 100\% in pneumonia patients.

The limitations of the HQCNN model in this work
are:
\begin{itemize}
\item The size of the COVID-19 image is used with 28x28
pixels which are divided into 4x4 pixel.
\item  The architecture of the HQCNN model is a small to
compatible with 4 qubits in the quantum simulator.
\item The used random quantum circuit works as preprocessing
layer for images.
\item The size of the used images in the training phase is
small.
\end{itemize}

\begin{table}[]
	\centering
	\caption{The HQCNN model performance with each class on D1, D2, and D3.}
	\label{Tbl:T10}\resizebox{.5\textwidth}{!}{%
		\begin{tabular}{cccccc}
			\hline
			\textbf{Class}	& \textbf{ACC.(\%)}	&\textbf{Sns.(\%)} &\textbf{Prc.(\%)}&\textbf{F1-measure(\%)} \\  \hline
			\multirow{4}{*}{}&  \hspace{3cm} D1&  \\ \hline
			normal & 97.8 & 100& 98 & 99 \\
			COVID-19 & 99.3 &  97& 99& 98 \\ \hline
			\multirow{4}{*}{}&  \hspace{3cm} D2&  \\ \hline
			COVID-19 & 97.5& 99 & 98 &98\\
			pneumonia &99.7 & 99& 100 &99 \\\hline
			\multirow{4}{*}{}&  \hspace{3cm} D3&  \\ \hline
			normal &89.3 & 77&  89 &83\\
			COVID-19 & 97.5& 99& 98 &98\\
			pneumonia &94.6 & 92& 85 &88 \\
			\hline
	\end{tabular}}
\end{table}

\section{Conclusion and Future Work}
\label{Sec:Con}
In this work, a new hybrid quantum-classical CNNs (HQCNN) model has been proposed for COVID-19 classification with chest radiography images. The aim of the HQCNN model combines the random quantum circuit with CNNs and to enhance the performance. The HQCNN model has been used the random quantum circuit as a quantum convolution layer to compute convolution operation on a quantum device. The HQCNN model has been evaluated on binary and multi-class COVID-19 datasets, it outperformed classical CNNs and various models. It achieved an accuracy of 99\% and 88.6\%, a sensitivity of 99.3\%, and 88.7\% for binary, and multi-class datasets, respectively. The future work will be highly focused on using different encoding methods (i.e., amplitude encoding) with further quantum convolutional layers. Also, we will enhance the
HQCNN architecture to overcome on limitations of the
proposed model.

\section*{Author Contribution statement}
Essam H. Houssein: Supervision, Methodology, Conceptualization, Formal analysis, Writing - review \& editing. Zainab Abohashima: Software,  Resources, Data Curation, Writing - original draft. Mohamed Elhoseny: Conceptualization, Formal analysis, Writing - review \& editing. Waleed M. Mohamed: Conceptualization, Formal analysis, Writing - review \& editing. All authors read and approved the final paper.

\section*{Conflict of interest}
The authors declare that there is no conflict of interest. Non-financial competing interests. 

\section*{Compliance with ethical standards}
This article does not contain any studies with human participants or animals performed by any of the authors.

\section*{Acknowledgments}
The authors would like to thank Pennylane's team members for helpful discussions, especially, Dr Andrea Mari for valuable discussions.

\bibliographystyle{unsrt}
\bibliography{refs}
%%
%%{\footnotesize
%%\bibliographystyle{icle-num}
%%}
%%% Non-BibTeX users please use
%\bibliography{refs}
%=====================================
%%%%%%%%%%%%%%%%%%%%%%%%%%%%%%%%%%%%%%%%%%
\end{document}